\documentclass[12pt,a4paper]{article}
\usepackage{amsmath,amsthm,amsfonts,amssymb,amscd}
\usepackage{graphicx}
\newtheorem{theorem}{Theorem}

\newcommand{\ov} {\overline}

\newcommand{\dt} {\delta}

\def\Z{{\mathbb Z}}

\def\R{{\mathbb R}}

\def\C{{\mathbb C}}

\newcommand{\bproof}{\noindent {\it Proof. }}
\newcommand{\eproof}{\hfill$\Box$}
\newcommand{\nd} {\noindent}

\begin{document}
\begin{titlepage}

  \nd
  Published in: Journal of Nonlinear Science (2022)\\ \nd
  https://doi.org/10.1007/s00332-022-09817-6
  \vskip 2truecm

\begin{center}

  {\bf Errata and Addenda to:  ``Hydrodynamic Vortex on Surfaces''  and ``The motion of a vortex on a
closed surface of constant
negative curvature''}
  \vskip 1truemm
{    \LARGE\bf }

\end{center}

\vskip  1.0truecm
\centerline {{\large Clodoaldo Grotta-Ragazzo$^*$.}}

\vskip 0.5truecm



\begin{abstract}
  The two papers in the title contain some equations that are not complete.
  The missing terms, which are of topological origin,
  were recently unveiled by Bj\"orn Gustafsson.
  In this note we present the equations of Gustafsson in the case of a single
  vortex in a compact boundaryless surface, and show that many conclusions
  we have reached  with the incomplete equations
  remain valid. It seems that the
  the extra-terms in
  Gustafsson's equations
  can be explicitly written in terms of elementary functions
  only for the two-torus, and this is done at the end of this note.
\end{abstract} 

\vskip .5truecm

\nd



\vfill
\hrule
\smallskip
\nd
$^*$ Instituto de Matem\'{a}tica e Estat\'{i}stica da
Universidade de S\~{a}o Paulo,\\
\nd Rua do Mat\~ao 1010, 05508-090, S\~{a}o Paulo, SP, Brazil.\\
\nd Partially supported by FAPESP grant 2016/25053-8.\\
 \nd email: ragazzo@usp.br \\
\nd ORCID: 0000-0002-4277-4173
\end{titlepage}

\pagebreak
\section{Gustafsson's equations.}

In a recent  paper Bj\"orn Gustafsson \cite{gustafsson2022vortex}
presented equations for
the motion of hydrodynamic vortices on closed surfaces (compact and boundaryless).
These equations, which will be referred as Gustafsson's  equations, are different
from those used by previous authors in the case of surfaces of genus $g$ greater
than zero. Gustafsson's introduced extra-terms that correctly
couple the motion of the
vortices to the variation of the harmonic part of the velocity field.

In this note  we present the equations of Gustafsson in the case of a
single vortex of intensity one in a closed surface of a  genus greater than
zero. We will
derive Gustafsson's equation not using a renormalized energy as done
in \cite{gustafsson2022vortex}, but our weak formulation of Euler's equation
given in \cite{ragvil}.
 The reader must consult \cite{gustafsson2022vortex} for
the Hamiltonian formulation of the equations of a system of vortices.
The ideas in \cite{gustafsson2022vortex}
can be used to correct the equations obtained in
\cite{ragvil} for systems of vortices in  surfaces with boundaries and/or ends
and genus greater than zero. 
Gustafsson's result do not
change the equations in \cite{ragvil}
when the surface has genus zero but
it is not simply connected due to the presence of boundaries or ends, and this
 we will be  explained below.

\begin{figure}[hptb!]
 \includegraphics[width=0.95\textwidth]{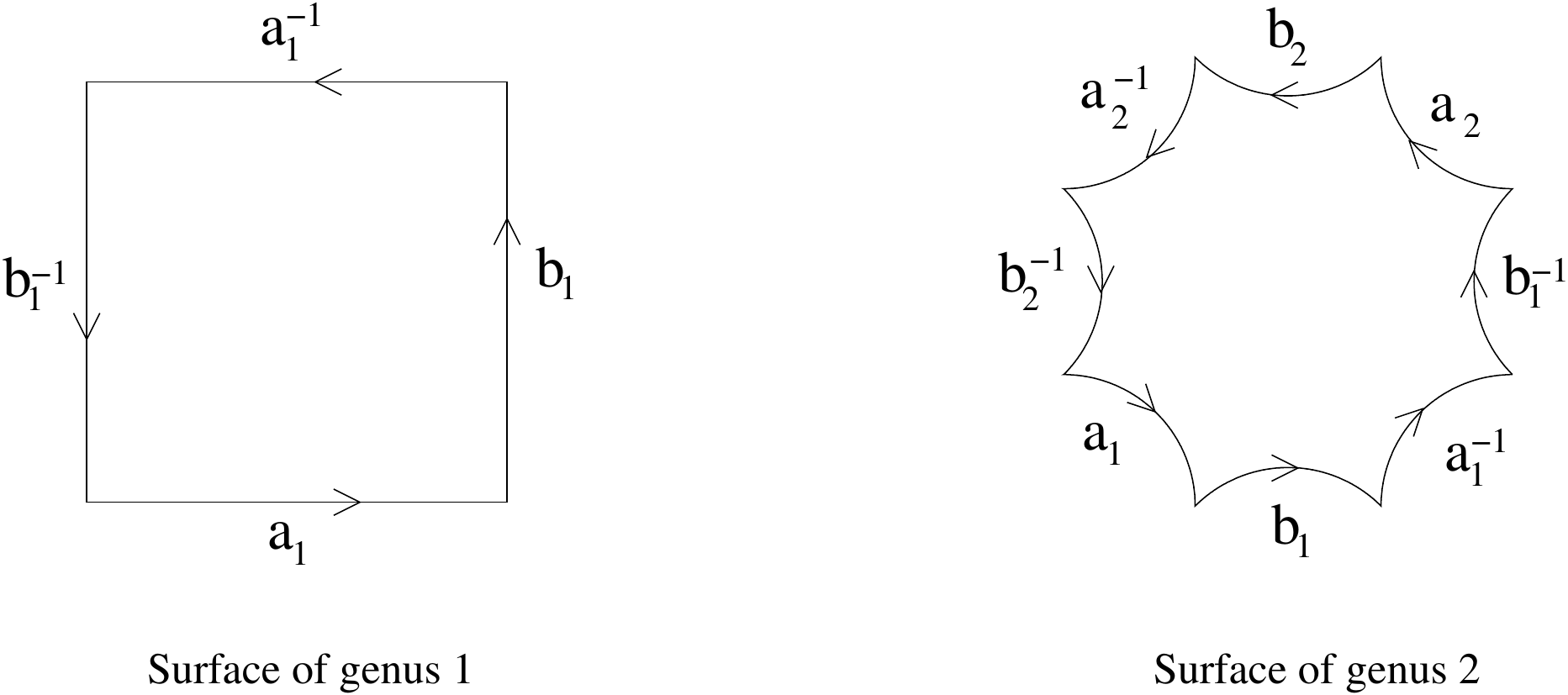}
\caption{Normal forms of surfaces of genus 1 and 2.}
\label{fig1}
\end{figure}

Every closed surface $S$  of genus $g\ge 1$ can
be obtained topologically by identifying in pairs appropriate
sides of a $4g$-sided polygon
  (\cite{farkas1992riemann} chapter 1).
In Figure \ref{fig1} we illustrate the construction of a torus (genus 1)
and of a bi-torus (genus 2).
Each  one of the sides  of the polygon
$a_1b_1a_1^{-1}b^{-1}_1\ldots a_gb_ga_g^{-1}b_g^{-1}$
corresponds to a curve
in a different homology class of the surface, 
and  the set of  $2g$ curves form a basis of  the homology of $S$. 
The intersection numbers of the curves
satisfy $a_j\cdot b_k=\delta_{jk}$, $a_j\cdot a_k=0$, and $b_j\cdot b_k=0$,
where $\delta_{jk}=1$ if $j=k$ and zero otherwise,
and there exist a basis
$\alpha_1,\ldots,\alpha_g,\beta_1,\ldots,\beta_g$ of  harmonic
1-forms such that
\begin{equation}
  \int_{a_k}\alpha_j=\delta_{jk}\,,\quad
  \int_{b_k}\alpha_j=0\,,\quad
   \int_{a_k}\beta_j=0\,,\quad
   \int_{b_k}\beta_j=\delta_{jk} \label{basis}
 \end{equation}
 and
 \begin{equation}
     \int_S\alpha_j\wedge\beta_k=\delta_{jk},\quad
     \int_S\alpha_j\wedge\alpha_k=\int_S\beta_j\wedge\beta_k=0\,,
     \label{basis2}
     \end{equation}
 for $j,k=1,\ldots,g$  (see \cite{farkas1992riemann} chapter 3).
 Note: the harmonic forms are conformal invariants.

 We are interested in the motion of an incompressible fluid of density one in
 a Riemannian surface $S$.
 The velocity field of the fluid will be denoted by $v$ and the corresponding
 velocity one-form by $u = \langle v,\cdot\rangle$,
 where $\langle \cdot,\cdot\rangle$
 is the inner product defined by the Riemannian metric.
 The Riemannian two-form of area will be  denoted by  $\mu$ and the
 vorticity two-form of the fluid  by $du$.
 Euler's equation for the
 motion of the fluid can be written as (\cite{ragvil} Eq. (3.3) after using Cartan's
 identity  $L_v u=i_vdu+d u(v)$):
 \begin{equation}
   \partial_t u+i_v du=-d\left( p+ \frac{|v|^2}{2}\right)\label{euler}
 \end{equation}

 A hydrodynamic vortex is  a singular velocity field  defined in the
 following way. 
 The Green's function of $S$ is the unique function that  satisfies
\begin{equation} 
-\Delta_q G(q,p)=\dt_p(q)-V^{-1}\, , \qquad \int_S G(q,p)\mu(q)=0\,,
\label{compacteq}
\end{equation} 
where $\Delta=\ast d\ast d$
is the Laplace-Beltrami operator\footnote{ If $\theta_1$ and $\theta_2$ are two orthonormal one-forms on an open set in $S$,
 then the Hodge-star operator ``$\ast$'' associated with the Riemannian 
metric is a linear map onto the space of differential forms that satisfies:
\[
 \ast 1=\theta_1\wedge\theta_2=\mu,\qquad 
\ast\theta_1=\theta_2,\qquad \ast\theta_2=-\theta_1, \qquad
 \ast\mu=1.
\]} on $S$, $\delta_p$ is the
$\delta$-distribution of Dirac, and $V$ is the volume of $S$.
A hydrodynamic vortex of intensity one at a point $p$   
   is the  velocity field $q\to v(q)$  associated with the velocity
form $-\ast d_qG(q,p)=\langle v(q),\cdot\rangle$.

The weak formulation of Euler's equation presented in Section 5 of
\cite{ragvil} implies that a hydrodynamic vortex at $p$ must
be carried by the regular part of the flow  (Eq. (5.7) in ibid.)
\begin{equation}
  \dot p= v_r(p)\,,\label{weak1}
\end{equation}
where $v_r(p)$ is the regular part of the velocity field $v$ at $p$.
The one-form $U=\langle v_r,\cdot\rangle$, which corresponds to $v_r$,
has two terms
\begin{equation}
 U=-\ast d \frac{R}{2} +\eta\,, \label{U1}
\end{equation}
where $R$ and $\eta$ depend on $p$.
The function $R$ in  the first term is the Robin function that is 
  the regular part of the Green's
  function at $p$:
    \begin{equation}
R(p)=
\lim_{\ell(q,p)\to 0} \left[
G(q,p)+\frac{1}{2\pi}\log \ell(q,p)\right]\,,
\label{robin}
\end{equation}
where $\ell(q,p)$ is the Riemannian distance between $q$ and $p$.
The second term $\eta$ corresponds to a harmonic velocity field,
which may  exist  in the absence of vortices and that
generates a ``background flow''.

Any  harmonic form on $S$ can be written as a linear combination
 of the  forms $\alpha_1,\ldots,\alpha_g,\beta_1,\ldots,\beta_g$:
  \begin{equation}
    \eta= \sum_{j=1}^g A_j \alpha_j+ B_j\beta_j\,.\label{eta1}
  \end{equation}
  The velocity one-form
  of the vortex with background flow is
  \begin{equation}
    q\to u=-\ast d_q G(q,p)+\eta(q)
  \end{equation}
  where
  $G$ is the Green's function in equation (\ref{compacteq}).Following
 \cite{gustafsson2022vortex},
  if $u$ satisfies Euler's equation (\ref{euler}), then the time derivative
  of the circulation along the curve $a_1$ must satisfy
 \begin{equation}\begin{split}
   & 
   \frac{d}{dt}\int_{a_1} u=  \int_{a_1}\partial_tu=\\ &
   -\int_{a_1}i_v du +\underbrace{\int_{a_1}d\left( p+ \frac{|u|^2}{2}\right)}_{=0}=
   \frac{1}{V}\int_{a_1}\underbrace{i_v\mu}_{\ast u}=
  \frac{1}{V}\int_{a_1}\ast \eta\,,
   \end{split}
   \label{euler3}
  \end{equation} 
  where we have assumed that $p\notin a_1$ and used equation (\ref{compacteq})
  to write $du=-~d~\ast~d~G=-\frac{1}{V}\mu$.  Similar equations
  must hold for other generators $a_2,\ldots,a_g,b_1,\ldots,b_g$ of the homology.

  The left hand side of equation (\ref{euler3}) can be simplified using the
  identities (\cite{gustafsson2022vortex}
  Section 3.2)
  \begin{equation}
    d_p\int_{a_j}\ast_q d_q G(q,p)=\beta_j\,,\quad
    d_p\int_{b_j}\ast_q d_q G(q,p)=-\alpha_j\,,\quad j=1,\ldots,g\,,
    \label{apbt}
    \end{equation}
    where the integration is with respect to $q$ and
    $\alpha_j$ and $\beta_j$ depend on $p$.
    
    The harmonic forms
    $\{\ast\alpha_1,\ldots,\ast\alpha_g,\ast\beta_1,\ldots \ast\beta_g\}$
    also form a basis of the cohomology of $S$, that is related to the basis
    $\{\alpha_1,\ldots,\alpha_g,\beta_1,\ldots \beta_g\}$ by means of the matrix
    (see  \cite{farkas1992riemann} chapter 3 or
    \cite{gustafsson2022vortex} Eqs. (5.20) and (5.21))
    \begin{equation}
      \left(\begin{array}{c}\ast \alpha\\\ast\beta\end{array}\right)=
      \left(\begin{array}{ll}-\mathbf{R} &\mathbf{P}\\-\mathbf{Q} & \mathbf{R}^T\end{array}\right)
       \left(\begin{array}{c} \alpha\\\beta\end{array}\right)\,,\label{matrix}
     \end{equation}
     where $\mathbf{P}>0$, $\mathbf{Q}>0$ and  $\mathbf{R}$ are  $g\times g$ square matrices given by
     \begin{equation}
       P_{jk}=\int_S\alpha_j\wedge\ast \alpha_k\,,\quad
       Q_{jk}=\int_S\beta_j\wedge\ast \beta_k\,,\quad
       R_{jk}=\int_S\alpha_j\wedge\ast \beta_k\,,\label{PQR}
     \end{equation}
     and $\mathbf{R}^T$ denotes the transpose of $\mathbf{R}$.

     From equation (\ref{euler3}), and its variations for other elements
     of the homology $a_2,\ldots,a_g,b_1,\ldots, b_g$,  and  equations
     (\ref{apbt}) and (\ref{matrix}), we obtain:
     \begin{equation}\begin{split}
         &\dot A_k -\beta_k [\dot p]=-\frac{1}{V}\sum_{j=1}^g A_j R_{jk}+B_j Q_{jk}\\
         &\dot B_k +\alpha_k [\dot p]=\frac{1}{V}\sum_{j=1}^g A_j P_{jk}+B_j R_{kj}\,,
     \quad k=1,\ldots,g\,.    \label{gust1}
       \end{split}
       \end{equation}

       Equations (\ref{gust1}) must be complemented by equations (\ref{weak1}),
       i.e. $\dot p= v_r(p)$, and (\ref{U1}), i.e,
       $ U=-\ast d \frac{R}{2} +\eta$, where
       $U=\langle v_r,\cdot\rangle$ is the ``vortex velocity form''.
       Using the  relation $i_{\dot p}\mu=i_{v_r}\mu=\ast U$ the equation
       for the velocity of the vortex can be written as
       \begin{equation}\begin{split}
           &  i_{\dot p}\, \mu= \ast U= d \frac{R}{2} +\ast \eta\,,
           \quad \text{where}\\
          & \ast\eta=\sum_{j=1}^gA_j(\ast \alpha_j)+B_j(\ast \beta_j)\,.
         \end{split}\label{gust2}
       \end{equation}
         Equations (\ref{gust1}) and (\ref{gust2}) are those 
         obtained by Gustafsson \cite{gustafsson2022vortex} for the
         motion of a  unitary vortex in a closed surface.

At this point it is instructive to show why Gustafsson's results do not
change the equations in \cite{ragvil} when the surface has genus zero but
it is not simply connected due to the presence of boundaries or ends.
The analysis will be done in the context of an example.

Let $\mathcal A$ be  the Euclidean annulus $\mathcal A:= \{1<|z|<2;z=x+iy\in \C\}$.
There is a unique  Green's function $F$ on $\mathcal A$
that satisfies $-\Delta_q F(q,p)=
\delta_p(q)$ and $F(q,p)=0$ for $q\in\partial \mathcal A$. Consider the
velocity one-form $u=-\ast d_q F(q,p)$. We will check whether $u$ defined
in this way can be a solution to Euler's equation.
As we did in equation (\ref{euler3}), we will compute the
time derivative of the circulation along each boundary
component $|z|=j$, $j=1,2$,  of $\mathcal A$ (orientation is positive when
the interior of $\cal A$ is on the left-hand side of the boundary component).
 We use that
$-d_q\ast_q d_q F(q,p)=0$ for $q\ne p$, because $F$ is harmonic, and
we obtain
\begin{equation}
  \frac{d}{dt}\int_{|z|=j}u=
  \int_{|z|=j}\partial_t u
  =-\int_{|z|=j}d\left( p+ \frac{|U|^2}{2}\right)=0\,.\label{euler2}
  \end{equation}
  In analogy to equation (\ref{apbt}),  it can be verified that the functions
  defined by
\begin{equation}
\omega_j(p)=-\int_{|z|=j}\ast d_q F(q,p)= \int_{|z|=j}u\qquad j=1,2
\label{harmeas}
\end{equation}
are harmonic on $\mathcal A$, and satisfy 
$\omega_1(p)=1$  for $|p|=1$,  $\omega_1(p)=0$  for $|p|=2$; and
$\omega_2(p)=1$  for $|p|=2$,  $\omega_2(p)=0$  for $|p|=1$ (in this case
$\omega_1(p)=\frac{2}{|p|}-1$ and $\omega_2(p)=-\frac{2}{|p|}+2$).
If $p$ is moving and
 $\frac{d}{dt} \omega_j(p)\ne 0$, then  equation (\ref{euler2}) is false,
namely circulation is not conserved at the boundaries.
In order to fix $F$ we add to it  a convenient combination of $\omega_1$ and
$\omega_2$ such that the circulations at the boundary components of $\mathcal A$
remain constant. If we denote the circulations at the boundaries $|z|=1$ and
$|z|=2$ by $c_1$ and $c_2$, respectively, then we can prescribe $c_1$ and $c_2$
arbitrarily provided that $c_1+c_2=1$ due to Stokes theorem.
For instance,  if $c_1=0$ and $c_2=1$, then the
``hydrodynamic Green's function'', which is  used to define a vortex, is 
\cite{lin1, lin} (or \cite{ragvil} Eq. (3.17))
\begin{equation}
  G(q,p)=F(q,p)+\frac{\omega_1(p)\omega_1(q)}{4\pi}\,.\label{green1}
  \end{equation}

  The difference between the loss of connectivity due to boundaries and
  due to genus is that in the former it always possible to find  a
  Green's function that
  satisfy  the circulation conditions imposed by Euler's equation,
  possibly by adding harmonic functions, while
  in the latter it is necessary to  add harmonic forms to fulfill
  the circulation conditions, since
  the Green's function  cannot be modified.
  The mistake in \cite{ragvil}  was to have neglected the role of 
harmonic forms in surfaces with nontrivial genus in substituting
differentials of harmonic functions in surfaces with boundaries.

The effect of  Gustafsson's equation upon
the paper \cite{ragvil} is restricted to  the definition of vortices on
surfaces of genus greater than zero, since  no example of this kind of surface
was treated in ibid.
         The main effect of  Gustafsson's equation is in my paper
         \cite{ragazzo2017motion}:
``The motion of a vortex on a closed surface of constant negative
curvature''. In this reference, I numerically computed the Robin function
on the Bolza surface (of constant curvature and of genus 2)
and using equation (\ref{gust2})  without the term $\ast \eta$
I concluded that a vortex in the Bolza surface moves, as previously conjectured
by Jair Koiller.
The vortex trajectories  in  Figures 7 and 8 in ibid. may not be correct 
(these figures show  the level sets of the Robin function), however the conclusion
that the vortex moves is still correct.
Indeed,  the following four  main conclusions obtained
in the papers  \cite{ragazzo2017motion} and 
\cite{ragvil}
{\it remain valid after  Gustafsson's result}:
\begin{itemize}
\item[(a)] A vortex in the Bolza surface, a surface constant of
  negative curvature -1
  and genus 2,
  moves.
\item[(b)] If $S$ is
  a closed hyperelliptic surface of genus $g\ge  2$ and constant curvature, then
  every
  Weierstrass point of S is an equilibrium of
  the  motion of a single vortex (Theorem 3.4 in
  \cite{ragazzo2017motion}).
\item[(c)] It is possible to construct vortex crystals in the Poincar\'e disc
  with vortices placed at the lift of the Weirestrass points to  the
  universal covering (see Section 5 of
  \cite{ragazzo2017motion} for details).
\item[(d)] A  ``steady vortex metric'' is  a Riemannian metric
  in which a single vortex does not move regardless of  its position
  \cite{ragvil}. A Riemannian metric is a steady vortex metric
  if, and only if, its Robin function is constant.
\end{itemize}

The four statements above are corollaries of the following theorem.
         \begin{theorem} If  $dR(\ov p)=0$ for a certain $\ov p\in S$, then
           $\ov p$ and $\eta=0$  ($A_1=\ldots=A_g=B_1=\ldots=B_g=0$)
           comprise an equilibrium of Gustafsson's equation.

           For  Gustafsson's equation,    a
           vortex remains at rest regardless of its initial position
           if, and only if, the Robin function $R$ is constant and
           $\eta$ is chosen initially equal to zero.
                   \end{theorem}

         \bproof The first claim in the theorem is immediate:
         If $d R(\ov p)=0$ for a certain $p$, then $\ov p$ and $\eta=0$
         comprise an equilibrium of equations
         (\ref{gust1}) and (\ref{gust2}).

         The velocity of a vortex is null regardless of its position if, and
         only if, the velocity form $U (p)$ is null
for every $p\in S$. 
         The $L_2(S)$ norm of both sides of equation (\ref{gust2}) is given by
         \begin{equation}
           \int_S U\wedge \ast U= \int_S d \frac{R}{2}\wedge\ast d \frac{R}{2} +
           \int_S \eta\wedge \ast \eta \label{Un}
         \end{equation}
         because $\int_S d \frac{R}{2}\wedge \eta=0$. For any one form
         $\sigma=\sigma_1 \theta_1+\sigma_2\theta_2$, where $\theta_1,\theta_2$
         are local orthonormal one-forms,
         $\sigma\wedge\ast \sigma= (\sigma_1^2+\sigma_2^2) \theta_1\wedge\theta_2
         = (\sigma_1^2+\sigma_2^2) \mu$. Therefore, the left-hand side of equation
         (\ref{Un}) is zero if, and only if, both terms in the right-hand side
         are zero.
         \eproof

         {\bf Phase space, Hamiltonian function, and symplectic form.}

         Let $H^1(S)$ denote the space of harmonic forms on $S$.
          The phase space of Gustafsson's
         equation is the Cartesian product $S\times H^1(S)$. 
         A point in the phase space is a pair $(p,\eta)$ where $p\in S$ is a
         possible position of the vortex  and
         $\eta\in H^1(S)$ is a possible harmonic velocity form. Let 
         $\{\alpha_{1q},\ldots,\alpha_{gq},\beta_{1q},\ldots,\beta_{gq}\}$ be the 
         basis of $H^1(S)$ that satisfies the relations (\ref{basis}).
         The index $q\in S$ indicates that the form does not act on vectors defined
         on the first factor of the phase space $S\times H^1(S)$.

         We define the Bergman kernel
         \begin{equation}
           B_{qp}=\sum_{j=1}^g\alpha_{jp}\beta_{jq}-\beta_{jp}\alpha_{jq}\,,
         \end{equation}
         where the index $p$ indicates that the form
         acts on the first factor of $S\times H^1(S)$.
         Using equation (\ref{basis2}) it is easy to verify that,
         for any harmonic form $\sigma_q=\sum_kA_k\alpha_{kq}+B_k\beta_{kq}$,
         \begin{equation}
           \int_{S_q}\sigma_q\wedge B_{qp} =\sigma_p=\sum_jA_j\alpha_{jp}+
           B_j\beta_{jp}\,.
           \label{B2}
         \end{equation}
         Using this notation and the  definition of $B$ we can rewrite Gustafsson's
         equations (\ref{gust1}) and (\ref{gust2}) in a concise way:
         \begin{eqnarray}
           & &  \dot\eta_q+i_{\dot p}B_{qp}=\frac{1}{V}\ast \eta_q\label{gust3}\\
           & & i_{\dot p}\, \mu_p= d \frac{R_p}{2} +\ast \eta_p\,,
               \label{gust4}
               \end{eqnarray}
               where $i_{\dot p}B_{qp}=
               \sum_{j=1}^g\alpha_{jp}[\dot p]\beta_{jq}-\beta_{jp}[\dot p]\alpha_{jq}$.
\vskip 2mm
               Let $H$ be the function 
               \begin{equation}
                 H(p,\eta)=\frac{R(p)}{2}+\frac{1}{2}
                 \int_{S_q}\eta\wedge \ast\eta\,,\label{H}
               \end{equation}
               defined on the phase space\footnote{
                 If $\eta=\sum_{j=1}^gA_j\alpha_j+B_j\beta_j$, then using equations
               (\ref{matrix}) and (\ref{PQR}) 
               \begin{equation}
                 \int_{S_q}\eta\wedge \ast\eta=\left(A^T,B^T\right)
                 \left(\begin{array}{ll} \mathbf{P} & \mathbf{R}\\ \mathbf{R}^T &
\mathbf{Q}\end{array}\right)
                 \left(\begin{array}{l} A \\ B\end{array}\right)
                 \end{equation}}.

                 The function $H$ is a first integral of Gustafsson's equation,
                 that is 
$\frac{d H}{dt}= \frac{dR[\dot p]}{2}+
\int_{S_q}\dot\eta\wedge \ast\eta=0$.
Indeed, equation (\ref{gust4}) implies $ \frac{dR[\dot p]}{2}=-\ast\eta_p[\dot p]$
and equation (\ref{gust3}) implies
\[\begin{split}
 & \dot\eta_q\wedge \ast \eta_q+i_{\dot p}B_{qp}\wedge \ast \eta_q=
  \frac{1}{V}\ast \eta_q\wedge \ast \eta_q=0\Rightarrow\\
&\int_{S_q}\dot\eta_q\wedge \ast \eta_q=-\int_{S_q}i_{\dot p}B_{qp}\wedge \ast \eta_q
= \ast \eta_p[\dot p]\,, \end{split}\]
where we have used equation  (\ref{B2}).

{\it Note}: if $\ov p$ is a local minimum of the Robin function, then
conservation of $H$ and the positivity
of $\int_{S_q}\eta\wedge \ast\eta$ implies that $(p,\eta)=(\ov p,0)$ is a stable
equilibrium.

Let $\Omega$ be the following two form defined on $S\times H^1(S)$:
\begin{equation}
  \Omega[(\dot p,\dot \eta),(P,N)]=\mu(\dot p,P)-
  V\int_{S_q}(\dot \eta_p+i_{\dot p}B_{qp})\wedge
  (N_q+i_{P}B_{qp})\label{form}
\end{equation}
where $(\dot p,\dot \eta)$ and $(P,N)$ is a pair of vectors at a point
$(p,\eta)\in S\times H^1(S)$ and the integration is with respect to $q$.
In  the coordinates  $\eta=\sum_{j=1}^gA_j\alpha_j+B_j\beta_j$, 
 $\Omega$ can be written as
\begin{equation}
  \Omega=\mu-V \sum_{i=1}^g\big(dA_j-\beta_{jp}\big)\wedge\big(dB_j+\alpha_{jp}\big)\,.
  \label{form2}
  \end{equation}
  The forms $\alpha_{jp}$ and $\beta_{jp}$ are closed, and so in any contractible
  open ball $\mathcal B$ in $S$ there exist functions $\xi_j(p)$
  and $\zeta_j(p)$ such that
  $\alpha_{jp}=d\xi_j$ and $\beta_{jp}=d\zeta_j$. The map 
  $\mathcal B\times H^1(S)\to \mathcal B\times H^1(S)$ given  by
  \[
    \big(p,(A,B)\big)\to \big(p,(\tilde A,\tilde B)\big)\ \text{where}\ 
  \tilde A_j=A_j-\zeta_j(p)\,, \ \tilde B_j=B_j+\xi_j(p)\,,  j=1,\ldots,g
\]
defines new coordinates on $\mathcal B\times H^1(S)$. In these coordinates,
which were used in \cite{gustafsson2022vortex},
\begin{equation}
  \Omega=\mu-V \sum_{i=1}^gd\tilde A_j\wedge d\tilde B_j\,,\label{form3}
  \end{equation}
and it becomes clear that 
 $\Omega$ is symplectic.

We will show that 
$(\dot p,\dot \eta)$ given in equations (\ref{gust3}) and (\ref{gust4})
can be obtained from the equality
\begin{equation}
  \Omega[(\dot p,\dot \eta),(P,N)]=dH[(P,N)]=
  \frac{ dR}{2}[P]+\int_{S_q}N_q\wedge \ast \eta_q\,,
\end{equation}
that is valid  for any vector $(P,N)$ at the point
$(p,\eta)\in S\times H^1(S)$.
  For $P=0$  \[
    - V\int_{S_q}(\dot \eta_q+i_{\dot p}B_{qp})\wedge N_q
    =\int_{S_q}N_q\wedge \ast \eta_q\,,\]
  and  equation  (\ref{gust3})
  must hold. For $N=0$
  \[
      \frac{ dR}{2}[P]=\mu(\dot p,P)-
     V\int_{S_q}\underbrace{(\dot \eta_p+i_{\dot p}B_{qp})}_{=\frac{1}{V}\ast \eta_q \ \
     (\ref{gust3})}\wedge
   i_{P}B_{qp}
 =\mu(\dot p,P)-
 i_{P}\underbrace{\int_{S_q}\ast \eta_q \wedge B_{qp}}_{=\ast \eta_p \ \ (\ref{B2})}\]
that implies equation (\ref{gust4}).

         {\bf The equations of motion of a single vortex in the 2-torus}.

         Every Riemannian two torus is conformally equivalent to a flat
         torus $\R^2 /(\vec a\,\Z+\vec b\,\Z)$ where
         $\vec a=a_x\partial_x+a_y\partial_y$ and
         $\vec b=b_x\partial_x+b_y\partial_y$
         are the generators of the torus lattice. The volume form
         of the metric is given by $\mu=\lambda^2 dx\wedge dy$.
         We assume that the volume of both
        the torus and the flat torus are  one, namely
         \begin{equation}
          V=\int_S\mu=1\quad\text{and}\quad   a_x b_y-a_y b_x=1\,.
           \end{equation}
           The generators of the homology of the torus are the curves
           $a:=\{t\, \vec a: t\in[0,1)\}$ and  $b:=\{t\,\vec b: t\in[0,1)\}$.
           The basis of harmonic one forms $\{\alpha,\beta\}$
           such that $\int_a\alpha=\int_b\beta=1$ and
           $\int_a\beta=\int_b\alpha=0$ is given by
           \begin{equation}
             \alpha=b_y dx-b_x dy\,,\qquad \beta=-a_y dx+a_x dy\,,\label{ab}
           \end{equation}
           which implies
            \begin{equation}
             \ast\alpha=b_x dx +b_y dy\,,\qquad \ast \beta=-a_x dx-a_y dy\,.
           \end{equation}

           Let $\dot x\partial_x+\dot y\partial_y$
           denote the velocity of the vortex,
           $R$ be the Robin function associated with the Riemannian
    metric\footnote{In this case, it can be shown that 
             the Robin function is given by
             \[
               R=\frac{1}{2\pi}\log \lambda+2 \phi+ const\,,
             \]
             where $\phi$ is the solution to the equation in
             $\R^2 /(\vec a\,\Z+\vec b\,\Z)$
             \[
               \frac{\partial^2 \phi}{\partial x^2}+
               \frac{\partial^2 \phi}{\partial y^2}=\lambda^2-1\,.
             \]}, and
           $\eta=A\alpha+B\beta$.
           In this case Gustafsson's  equation  (\ref{gust2})
becomes
            \begin{equation}
              \begin{split}
                &\lambda^2\, \dot x= \ \, \frac{1}{2}\frac{\partial R}{\partial y}
                +b_y A-a_y B
                \\ &
                \lambda^2\, \dot y= -\frac{1}{2}\frac{\partial R}{\partial x}
                -b_x A+a_x B\,,
\end{split}\label{t1}\end{equation}
and equation (\ref{gust1}) can be written as
\begin{equation}\begin{split}
& \ \ \, \ b_y \dot A  -a_y\dot B -\dot y= b_x A- a_x B\\   
&-b_x \dot A+a_x \dot B+\dot x=b_y A-a_y B\,.
\end{split}\label{t2}\end{equation}

The Hamiltonian function as defined in equation (\ref{H}) is
\begin{equation}
 H=\frac{R(x,y)}{2}+\frac{1}{2} \big(A,B\big)\left(
\begin{array}{cc}
 b_x^2+b_y^2 & -a_xb_x-a_yb_y \\
 -a_xb_x-a_yb_y & a_x^2+a_y^2 \\
\end{array}
\right)\left(\begin{array}{c} A \\ B\end{array}\right)
\end{equation}

The symplectic form as defined in (\ref{form2}) is
\begin{equation}
   \Omega=\lambda^2 dx\wedge dy -(dA-\beta)\wedge(dB+\alpha)\,,
  \end{equation}
where $\alpha$ and $\beta$ are given in equation (\ref{ab}).

If the torus is flat, then $\lambda=1$, $R=$constant, and  equations
(\ref{t1}) and (\ref{t2}) reduce to
 \begin{equation}
           \dot x=b_y A-a_y B\,,\quad
                \dot y=  -b_x A+a_x B\,,\quad
                 \dot A=0\,,\quad \dot B=0\,.
\end{equation}
The vortex moves along straight lines following the harmonic part of
velocity field.

\vskip 1cm

\nd
Data sharing not applicable to this article as no datasets were
generated or analysed during the current study.
\bibliographystyle{plain}

\end{document}